\documentclass[preprint,showpacs,preprintnumbers,amsmath,amssymb,epsf,
tightenlines]{revtex4}

\usepackage{graphicx}
\usepackage{subfigure}

\begin{document}

\preprint{KAIST-TH 2003/10}

\title{Dipion invariant mass spectrum \\ in 
$X(3872) \rightarrow J/\psi \pi \pi$ }



\author{Taewon Kim}
\email[]{citadel@muon.kaist.ac.kr}%]%{Your e-mail address}
%\thanks{}
\affiliation{Department of Physics, KAIST \\ Daejon 305-701, Korea}

\author{P. Ko}
\email[]{pko@muon.kaist.ac.kr}%]%{Your e-mail address}
%\thanks{}
\affiliation{Department of Physics, KAIST \\ Daejon 305-701, Korea}


\date{\today}

\begin{abstract}
%We point out that The 
It is pointed out that dipion invariant mass spectrum in 
$X (3872) \rightarrow \pi \pi J/\psi$ is a useful probe for the $J^{PC}$ 
quantum number of the $X(3872)$, complementary to the angular distributions. 
If $X(3872)$ is a $^1 P_1$ state, the dipion has a peak at low $m_{\pi\pi}$ 
region, which is not in accord with the preliminary Belle data. If 
$X(3872)$ is a $^3 D_2$ or $^3 D_3$ state, the dipion spectrum shows a peak 
at high $m_{\pi\pi}$ region, which is broader than the $\rho$ resonance that
might come from the decay of a molecular state: $(D \overline{D^*}) (I=1) 
\rightarrow J/\psi \rho \rightarrow J/\psi \pi^+ \pi^-$.
Better measurement of $m_{\pi\pi}$ spectrum will shed light on the nature
of $X(3872)$. 
\end{abstract}

\pacs{}
%\keywords{}
\maketitle

\section{Introduction}
%\subsection{}
%\subsubsection{}

%Some years ago, it was suggested that the 
$B$ decay is a nice place to look for charmonium states which are either 
above the $D^{(*)} \bar{D^{(*)}}$ threshold or have quantum numbers 
that are not accessible from $n ^3S_1$ states by cascade decays.
$2 ^3P_J$,$^3D_J$ and $1 ^1P_1$ states are such examples 
\cite{ko,ko2,eichten}. 
In this regard, the advance in NRQCD enabled us to estimate the inclusive 
$B$ decays into charmonia without infrared divergence problem \cite{bbl}. 

Recently, Belle collaboration reported a new narrow resonance $X(3872)$ in
\[
B \rightarrow X K \rightarrow ( J/\psi \pi \pi ) K
\]
decay channel \cite{skchoi}, which was subsequently confirmed by CDF 
Collaboration \cite{Acosta:2003zx}.  It is important to identify the nature 
of this new narrow state, and there are several works on this state already 
\cite{suzuki,Voloshin:2003nt,Tornqvist:2003na,Close:2003sg,Yuan:2003yz,
Wong:2003xk,Braaten:2003he,Barnes:2003vb,Swanson:2003tb,Eichten:2004uh,
Braaten:2004rn,Braaten:2004fk,Rosner:2004ac}. In particular, 
Pakvasa and Suzuki pinned down 
possible quantum numbers for $J^{PC} (X)$ \cite{suzuki} as follows: 
\begin{itemize} 
\item If it is a charmonium, it should be either $1 ^3D_2 (2^{--})$ or
$2 ^1 P_1  ( 1^{+-} )$  
\footnote{
We use the spectroscopy notation, $n ^{2S+1}L_J$ where $n$ is the
principal quantum number, and the $J^{PC}$ quantum number is shown in the 
parenthesis.}. 
Also $1 ^3D_3 (3^{--})$ is another possibility \cite{private}.
\item If $X(3872)$ is the $D \overline{D^*}$ molecular state, it should be
either $J^{PC} = 1^{+-}$ with $I=0$, or $J^{PC} = 1^{++}$ with $I=1$.
\end{itemize}
Also they pointed out that the dipion angular spectrum 
could be useful to determine  the $J^{PC}$ quantum number of 
$X (3872)$ \cite{suzuki}. 

In this letter, we show that the dipion invariant mass spectrum in 
$X  \rightarrow J/\psi \pi \pi $ provides 
independent information on the nature of $X(3872)$. Our method is 
complementary to the angular correlations suggested in Ref.~\cite{suzuki}, 
and is already useful to eliminate a possibility that $X = {^1P}_1$ state. 

%%%%%%%%%%%%%%%%%%%%%%%%%%%%%%%%%%%%%%%%%%%%%%%%%%%%%%%%%%%%%%%%%%%%%%%%%%%
\section{Chiral lagrangian involving heavy quarkonia and light verctor 
mesons}

Hadronic transition between heavy quarkonia can be described in terms of
QCD multipole expansion or chiral perturbation theory. 
Since the allowed dipion invariant mass in $X (k , \epsilon_X ) \rightarrow 
J/\psi (k^{'}, \epsilon_\psi ) ~\pi ( p_1 ) \pi ( p_2 )$ is 
\[
2 m_\pi \leq m_{\pi\pi} \leq ( M_X - M_{J/\psi} ) = 775~{\rm GeV}~
\approx m_\rho, 
\]
both approaches will be  suitable to our purpose. 
In this letter, we use chiral lagrangian approach, since we are interested 
only in the spectrum and not in the absolute decay rate. 
Under global chiral $SU(3)_L \times SU(3)_R$ transformation, 
the pion field $\Sigma(x) \equiv exp( 2 i \pi (x)/f_\pi )$ transforms as
\begin{equation}
\Sigma (x) \rightarrow L \Sigma (x) R^{\dagger}. 
\end{equation} 
Under parity $P$ and charge conjugation $C$, the pion fields transform as
\begin{eqnarray}
P: & \pi (t, \vec{x}) \rightarrow & - \pi (t, -\vec{x}) 
\nonumber  \\
   & \Sigma   (t, \vec{x}) \rightarrow & \Sigma^{\dagger} ( t , -\vec{x}) 
\nonumber  \\ 
C: & \pi (t, \vec{x}) \rightarrow &  \pi (t, \vec{x})^T
\nonumber  \\
   & \Sigma   (t, \vec{x}) \rightarrow & \Sigma (t, \vec{x})^T 
\end{eqnarray}

In order to introduce other matters such as $\rho$ or $X(3872)$ etc., 
it is convenient to define another field $\xi (x)$ by $\Sigma ( x ) 
\equiv \xi^2 (x) $, which transforms as
\begin{equation}
\xi (x) \rightarrow L \xi (x) U^{\dagger} (x) = U(x) \xi (x) R^{\dagger} .
\end{equation} 
The $3\times 3$ matrix field $U(x)$ depends on Goldstone fields $\pi (x)$ 
as well as the $\textrm{SU}(3)$ transformation matrices $L$ and $R$. 
It is convenient to define two vector fields with following properties
under chiral transformations:
\begin{eqnarray}
V_{\mu} = 
\mbox{$\frac{1}{2}$}
\,( \xi^\dagger \partial_\mu \xi + \xi \partial_\mu  \xi^\dagger ), 
&& V_\mu \to  U V_\mu U^{\dagger}  + U \partial_\mu U^{\dagger},
\nonumber\\
A_{\mu} =
\mbox{$\frac{i}{2}$}
\,( \xi^\dagger \partial_\mu \xi - \xi \partial_\mu \xi^\dagger),
&&A_\mu \to  U A_\mu U^{\dagger}.
\end{eqnarray}
Note that $V_\mu$ transforms like a gauge field. 

Chiral symmetry is explicitly broken by non-vanishing current-quark masses
and electromagnetic interactions. The former can be included by regarding 
the quark-mass matrix $m = \textrm{diag}\,( m_u , m_d , m_s )$ as a spurion 
with transformation property 
$m \rightarrow L m R^{\dagger} = R m L^{\dagger}$. 
It is more convenient to use $\xi m \xi + \xi^{\dagger} m \xi^{\dagger}$,
which transforms as an $\textrm{SU}(3)$ octet:
\begin{equation}
( \xi m \xi + \xi^{\dagger} m \xi^{\dagger} ) \rightarrow
U(x) ( \xi m \xi + \xi^{\dagger} m \xi^{\dagger} ) U^\dagger (x)
\end{equation}

One can  introduce light vector mesons $\rho_{\mu}$, which transforms as
\begin{equation}
\rho_\mu (x) \rightarrow U (x) \rho_\mu (x) U^{\dagger} (x) + 
U (x) \partial_\mu U^{\dagger} (x),
\end{equation}
under global chiral transformations \cite{Bando:1985rf}. 
Then $\rho_\mu (x)$ transforms as a gauge field under local 
$\textrm{SU}(3)$'s defined by Eq.~(1), as $V_\mu$ does. 
The covariant derivative $\mathcal{D}_\mu$ can be defined using $\rho_\mu$
instead of $V_\mu$. 
Note that $( \rho_{\mu} - V_\mu )$ has a simple transformation property  
under chiral transformation: 
\begin{equation}
 (\rho_{\mu} - V_\mu ) \rightarrow U(x)  
 (\rho_{\mu} - V_\mu )
U^{\dagger} (x), 
\end{equation}
Both $\rho_\mu$ and $V_\mu$ are $C-$ and $P-$odd. % $J^{PC} = 1^{-}$. 
Then it is straightforward to construct chiral invariant lagrangian using 
$\rho$ fields. In terms of a field stringth tensor $\rho_{\mu\nu}$,
%For example, 
\begin{equation}
{\cal L}_{\rho} = -\frac{1}{2}~ {\rm Tr} ( \rho_{\mu\nu} \rho^{\mu\nu} ) 
+ \frac{1}{2}~m_{\rho}^2 ~ {\rm Tr} ( \rho_{\mu} - V_\mu )^2 
%\\
%& + & \alpha \left[ \overline{\mathcal{P}} 
%         ( \rho \!\!\!/ - V \!\!\!\!\!/ \, ) B 
% + \overline{B} 
%    ( \rho \!\!\!/ - V \!\!\!\!\!/ \, ) {\mathcal{P}}  \right] 
%+ ...
\end{equation}
The 1st term is the kinetic term, and the second term is the $\rho$ mass 
term.

Since the final charmonium is moving very slowly in the rest frame of the
initial state $X(3872)$, we can use the heavy particle effective 
theory approach, by introducing a velocity dependent field
$X_v (x) \equiv X e^{i m_X v\cdot x}$, and similarly for $J/\psi$ field 
$\psi_v (x)$ \cite{casalbuoni}. 
If $X_v$ is an isosinglet, it is chially invariant. If $X_v$ is an isovector 
$I( X_v ) = 1$, then  it transforms as   
\begin{equation}
X_v (x) \rightarrow U(x) X_v (x) U^\dagger (x).
\end{equation}
This will be useful when we consider the case $J^{PC} (X) = 1^{++}, I=1$ 
in Sec.~VI.  

Then we can construct chiral lagrangian which is invariant under parity $P$
and charge conjugation $C$, using the pion field $U (x)$,   
$X_v (x)$, a chiral singlet $\psi_v (x)$, the metric tensor $g_{\mu\nu}$ and 
the Levi-Civita tensor $\epsilon_{\mu\nu\alpha\beta}$. 
Transformation properties of $X_v (x) $, $\psi_v (x)$ and $v^{\mu}$ under 
parity and charge conjugation are given in Table~1.

 \begin{table}
 \caption{\label{table1} 
Transformation properties of $X_v$, $\psi_v$ and $v$ 
under parity and charge conjugation. }
 \begin{ruledtabular}
 \begin{tabular}{ccc}
 Fields & $P$ & $C$ 
 \\   \tableline
 $v^{\mu}$ & $v_\mu = ( v^0 , - \vec{v} )$ & $v^{\mu} = ( v^0 , \vec{v} )$ 
 \\
 $\psi_v^{\mu}$  & $ \psi_{v\mu}$    & $ - \psi_v^\mu$ 
 \\
 $X_v^{\mu\nu} (^3 D_2 )$ & $-X_{v\mu\nu} $  & $ - X_v^{\mu\nu}$
 \\
 $X_v^{\mu\alpha\beta} (^3 D_3 )$ & $+ X_{v\mu\alpha\beta} $  
 & $ - X_v^{\mu\alpha\beta}$
 \\
 $X_v^\mu (^1 P_1 )$      & $ - X_{v\mu}$  & $- X_v^\mu $ 
 \\
 $X_v^\mu ( J^{PC} = 1^{++} )$ & $ - X_{v\mu} $ & $X_v^\mu $ 
 \end{tabular}
 \end{ruledtabular}
 \end{table}
%In the following, we use 
Using our chiral lagrangian, we write down the amplitude for 
$X \rightarrow J/\psi \pi \pi$ and derive the dipion mass spectrum. 
This approach gives the same results as QCD multipole explansion for the 
Lorentz and chiral structures of the amplitude,  upto the unknown overall 
normalization. Similar analyses for 
$\Upsilon (3S) \rightarrow \Upsilon (1S) \pi \pi$ were presented in 
Ref.s~\cite{koups3}. 

A remark is in order before we proceed. 
For  $\psi^{'} \rightarrow J/\psi \pi \pi$ or
$\Upsilon (3S) \rightarrow \Upsilon (1S) \pi \pi$, there are 4 independent
chirally invariant operators to lowest order in pion momenta:
\begin{equation}
{\cal M} \sim \epsilon\cdot\epsilon^{'} \left( q^2 + B E_1 E_2 + C m_\pi^2 
\right) + D \left( \epsilon\cdot p_1 \epsilon^{'} \cdot p_2 + 
  \epsilon\cdot p_2 \epsilon^{'} \cdot p_1 \right),
\end{equation}
with $B,C,D \sim O(1)$, but QCD multipole expansion predicts that $B=D=0$.
$B,C,D$ being apriori unknown, one cannot make definite predictions for 
the $m_{\pi\pi}$ spectrum. Rather one has to fix $B,C,D$ from the 
$m_{\pi\pi}$ spectrum as done in Ref.s~\cite{koups3}. On the other hand, 
there is only single chirally invaraint operator to lowest order in pion 
momenta in the three options for the $J^{PC} (X)$ we consider in this work,
and we can make a definite prediction for $m_{\pi\pi}$ sprectrum, unlike the
case for $\psi^{'} \rightarrow J/\psi \pi \pi$ or
$\Upsilon (3S) \rightarrow \Upsilon (1S) \pi \pi$.  

%%%%%%%%%%%%%%%%%%%%%%%%%%%%%%%%%%%%%%%%%%%%%%%%%%%%%%%%%%%%%%%%%%%%%%%%%%%%%
\section{$ 1 ^3D_2 (2^{--}) \rightarrow J/\psi \pi \pi $}

This decay is described by the following chiral lagrangian with heavy 
quarkonia:
\begin{equation}
{\cal L} = g (^3 D_2 )~\epsilon^{\mu\nu\alpha\beta}~v_\mu \psi_{v,\nu}
X_{v,\alpha\rho}~{\rm Tr}~\left[ \partial_\beta \Sigma 
\partial^\rho \Sigma^{\dagger}
\right] + h.c. 
\end{equation}
where the coupling $g ( ^3D_2 )$ is unknown parameter, that should be 
determined by the data or could be calculated within QCD multipole expansion.
(See Ref.~\cite{Moxhay:1987ch} for an explicit form of $g ( ^3D_2 )$.)
Then the amplitude for this decay is given by 
\begin{equation}
{\cal M} \sim \epsilon_{ijk} ~\epsilon^i_\psi \epsilon^{jl}_X 
~\left( p^k p^l - q^k q^l \right), 
\end{equation}
where $p \equiv p_1 - p_2$ and $q \equiv p_1 + p_2$, and $i,j,k,l=1,2,3$ 
are space indices.  
Note that there is only one operator that contributes to the decay
$ 1 ^3D_2 (2^{--}) \rightarrow J/\psi \pi \pi$, so that we can predict
the $\pi\pi$ spectrum without any ambiguity. The overall normalization 
is determined by the decay rate, but is irrelevant to our discussion for 
the $\pi\pi$ spectrum. 
Now it is straightforward to calculate the dipion invariant mass spectrum
using the above amplitude.   
The resulting spectrum is shown in Fig.~1 in the (red) solid curve. 
Note that the dipion invariant mass has a peak at high $m_{\pi\pi}$ region, 
which is consistent with the preliminary Belle data \cite{skchoi}.

%%%%%%%%%%%%%%%%%%%%%%%%%%%%%%%%%%%%%%%%%%%%%%%%%%%%%%%%%%%%%%%%%%%%%%%%%%%
%%  New Section on 3D3 (3^{--}) 
%%%%%%%%%%%%%%%%%%%%%%%%%%%%%%%%%%%%%%%%%%%%%%%%%%%%%%%%%%%%%%%%%%%%%%%%%%%
\section{${^3 D}_3 ( 3^{--} ) \rightarrow J/\psi \pi \pi$}

Another possibility for $X(3872)$ is that it is a ${^3 D}_3 ( 3^{--} )$ 
state, whose transformation properties under parity and charge conjugation 
is listed in Table~1. In this case again, there is a single chirally 
invariant operator that contributes to $X \rightarrow J/\psi \pi\pi$:
\begin{equation}
{\cal L} = g(^3 D_3 )~\psi_{v \mu} X^{\mu}_{v \alpha \beta}~ 
{\rm Tr} \left[ \partial^\alpha \Sigma 
\partial^\beta \Sigma^{\dagger} \right], 
\end{equation}
which leads to the following amplitude:
\begin{equation}
{\cal M} \sim \epsilon_\mu \epsilon_X^{\mu \alpha \beta} 
\left( p_{1 \alpha} p_{2 \beta} + p_{2 \alpha} p_{1 \beta} \right).
\end{equation}
%Here the totally symmetric rank--3 tensor 
%$\epsilon_X^{\mu \alpha \beta} $ is the spin-3 polarization tensor,
%satisfying the following relations:
%\begin{eqnarray}
%v \cdot \epsilon_X^{\mu \alpha \beta} & = & 
%\epsilon^{\mu \alpha}_{X \alpha} = 0 ,
%\nonumber  \\
%\sum_{s} \epsilon_{\mu\alpha\beta} \epsilon_{\mu^{'} \alpha^{'} \beta^{'}} 
%& = & 
%\end{eqnarray}
Then we can predict the $m_{\pi\pi}$ spectrum without ambuguity as before.
The spin averaged squared amplitude for $^3D_3 \rightarrow J/\psi \pi\pi$ 
is equal to that for ${^3D}_2 \rightarrow J/\psi \pi\pi$, which was
also discussed within QCD multipole expansion in 
Ref.~\cite{Moxhay:1987ch}.  The resulting 
$m_{\pi\pi}$ spectra are the same as shown in the (red) solid curve in 
Fig.~1. 

%%%%%%%%%%%%%%%%%%%%%%%%%%%%%%%%%%%%%%%%%%%%%%%%%%%%%%%%%%%%%%%%%%%%%%%%%%%

\section{$X ( 1^{+-}, I = 0 ) \rightarrow J/\psi \pi \pi$}

If $X$ is a charmonium  $^1 P_1$ state with $J^{PC} = 1^{+-}$ and $I=0$, 
the relevant chiral lagrangian is given  by 
\begin{equation}
{\cal L} = g (^1 P_1 )~\epsilon^{\mu\nu\alpha\beta}~
X_{v \mu} \psi_{v \nu}~{\rm Tr}~\left[ 
\partial_\alpha \Sigma \partial_\beta \Sigma^{\dagger} \right] 
\end{equation}
where the coupling $g (^1 P_1 )$ is an unknown parameter, and could be 
determined by the data or could be calculated within QCD multipole expansion
\cite{Voloshin:em,Ko:1992fk}.
Then, the amplitude for $X ( 1^{+-}, I = 0 ) \rightarrow J/\psi \pi \pi$ 
is given by 
\begin{equation}
{\cal M} \sim \epsilon_{ijk} \epsilon^i_X \epsilon^j_{\psi} 
~\left( E_1 p_2^k + E_2 p_1^k \right) ,
\label{eq:1p1} 
\end{equation}
where $E_1$ and $\vec{p}_1$ are the energy and the three momentum of the 
pion 1, and similarly for the pion 2. As before, we have ignored the overall 
normalization, which is irrelevant to our discussion for the spectrum. 

If $X$ is a $D \overline{D^*}$ molecular state with $J^{PC} = 1^{+-}$, 
we cannot apply QCD multipole expansion. Still the chiral lagrangian 
approach will be applicable, and the above amplitude is still  valid. 

Using the above amplitude (\ref{eq:1p1}), it is straightforward to 
calculate the $\pi\pi$ invariant mass spectrum for 
$X ( 1^{+-} ) \rightarrow \pi \pi J/\psi$. 
We show the result in Fig.~1 in the (green) dotted curve. 
Note that the dipion spectrum has a peak at low $m_{\pi\pi}$ region, if 
$X$ has $J^{PC} = 1^{+-}$, which is  disfavored by the preliminary 
Belle data \cite{skchoi}. Once high statistics data is obtained, one can 
easily check if $J^{PC}(X) = 1^{+-}$ is a correct assignment or not.

%%%%%%%%%%%%%%%%%%%%%%%%%%%%%%%%%%%%%%%%%%%%%%%%%%%%%%%%%%%%%%%%%%%%%%%%%
\section{$X ( 1^{++}, I=1)  \rightarrow J/\psi \pi \pi$}

This case includes that the decaying state is $I=1$ $D \overline{D^*}$ 
molecular state and the final dipion is in $I=1$, which would be dominated
by $\rho$ meson. Since $\rho^0 \rightarrow \pi^0 \pi^0$ is forbidden by
angular momentum conservation and Bose symmetry, the $\pi\pi$ in $X 
\rightarrow J/\psi \pi\pi$ should be charged pions in this case.

Since charge-conjugation symmetry C is conserved in strong interaction, 
we must have $C(\pi\pi) = -1$, if $C(X) = +$. Now let us construct 
a $C-$even chiral invariant operator. Since  $I(X) = 1$ in this case, 
the $X$ field transforms as $X_v \rightarrow U(x) X_v (x) U^{\dagger} (x)$
as discussed in Sec.~II.
Therefore the lowest order chiral lagrangian for 
$X ( 1^{++}, I=1)  \rightarrow J/\psi \pi \pi$ is given by 
\begin{equation}
{\cal L} \sim \epsilon^{\mu\nu\alpha\beta}~v_\mu 
{\rm Tr} [ ( \rho_\nu - V_\nu ) X_{v \alpha} ]~ \psi_{v \beta}. 
\end{equation}
Then, the amplitude for $X ( 1^{++}, I=1)  \rightarrow J/\psi \pi \pi$  
is  given by
\begin{equation}
{\cal M} \sim \epsilon_{\mu\nu\alpha\beta}~v^\mu p^\nu \epsilon_X^\alpha 
\epsilon_\psi^\beta~\left[ 1 + { m_\rho^2 \over {q^2 - m_\rho^2 + i m_\rho 
\Gamma_\rho }} \right],
\end{equation}
where $q = p_1 + p_2$ and $p = p_1 - p_2$ as before, and 
$\Gamma_\rho = 153$ MeV is the total decay width of the $\rho$ meson.
The first term is from the contact interaction [ Fig.~1 (a) ], 
whereas the second term is due to the $\rho$ exchange [ Fig.~1 (b) ].

%The simplest nontrivial $C-$odd and chirally invariant operator will involve 
%\[
%{\rm Tr} [ ( \rho_\mu - V_\mu ) A_\nu ]. 
%\]
%Therefore $X( 1^{++} ) \rightarrow J/\psi + 3 \pi )$ should have been observed
%instead of $X( 1^{++} ) \rightarrow J/\psi  \pi \pi)$ ,
%if $X (3872)$ were C-even. Note that chiral symmetry and charge conjugation 
%(C) symmetry played an essential role to exclude the possibility 
%$J^{PC} (X) = 1^{++}$. In the earlier analyses, only isospin symmetry and 
%C symmetry were considered so that $J^{PC} (X) = 1^{++}$ option was viable,
%but this is no longer possible once we consider chiral symmetry properly. 
%Including chiral symmetry and C-symmetry exclude this possiblity, however. 

\begin{figure}
% \centerline{\epsfxsize=10cm \epsfbox{}}
  \begin{center}
    \includegraphics[width=6cm]{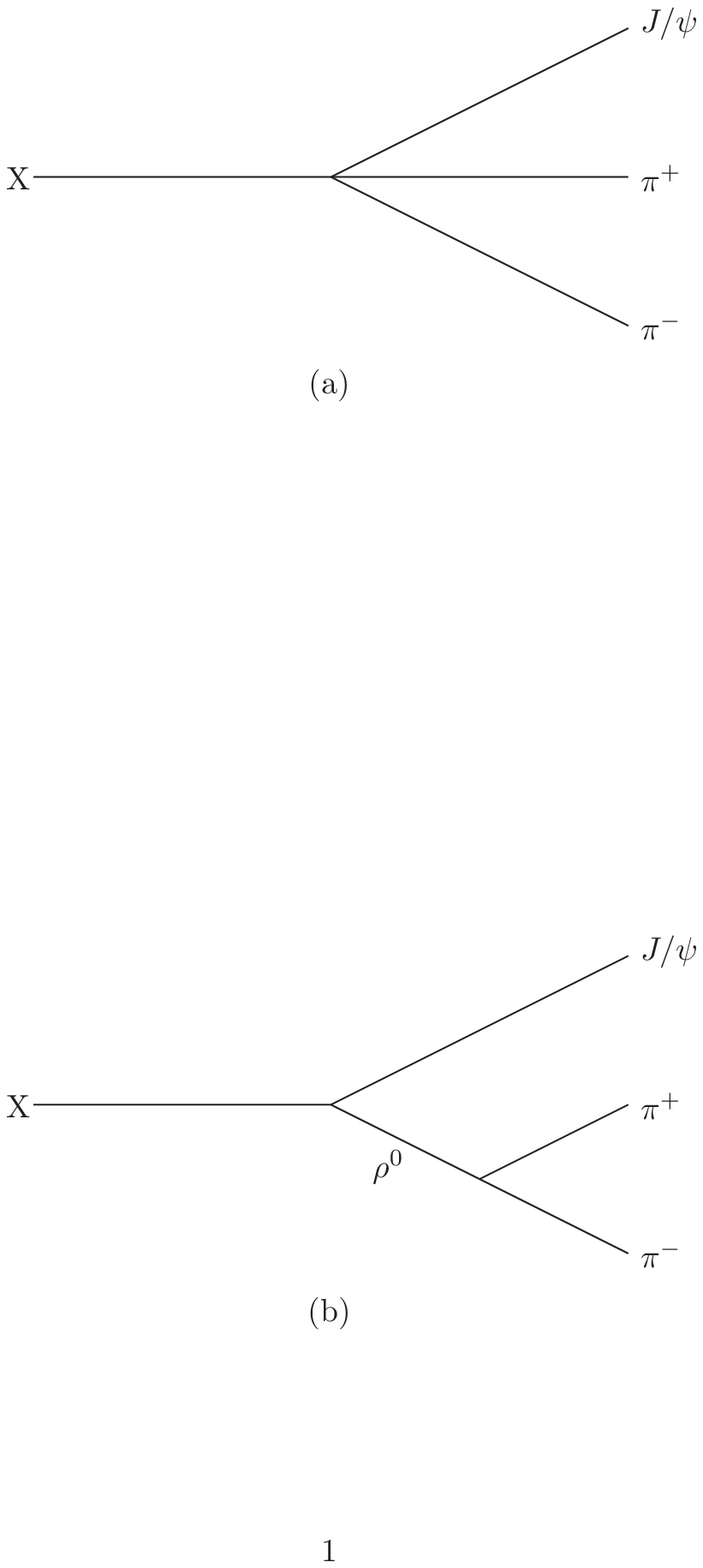}
  \end{center}
\caption{Feynman diagrams for $X(3872) \rightarrow J/\psi \pi\pi$
  }
\label{fig1}
\end{figure}

\begin{figure}
% \centerline{\epsfxsize=10cm \epsfbox{}}
  \begin{center}
    \includegraphics[width=12cm]{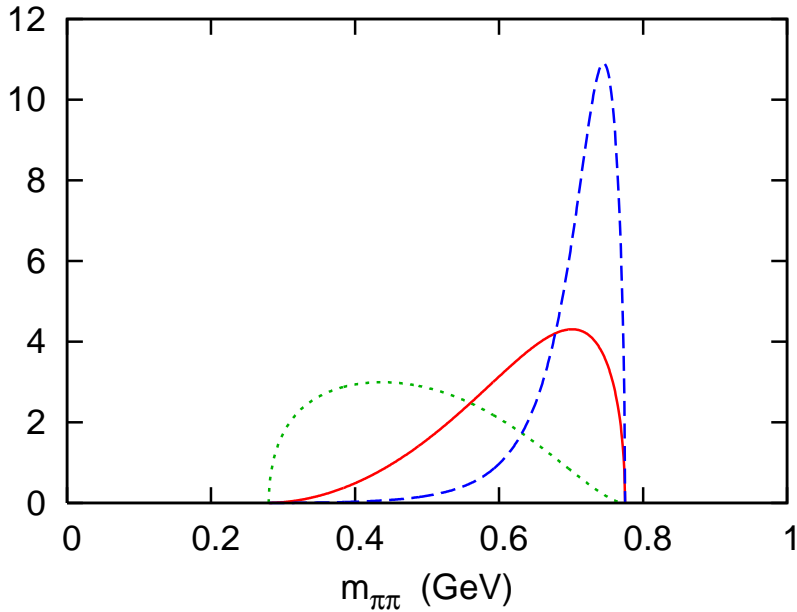}
  \end{center}
\caption{Dipion invariant mass spectra for $J^{PC} (X) = 2^{--}$ or $3^{--}$ 
(red and solid), $J^{PC} (X) = 1^{+-}$ (green and dotted) and 
$J^{PC} (X) = 1^{++}$ (blue and dashed).  The vertical scale is arbitrary 
for each case,  and the areas under the curves  is all the same. 
  }
\label{fig2}
\end{figure}
In this case, the polarization of $\rho$ and $J/\psi$ tends to 
be perpendicular with each other, which can be tested 
by measuring the three--momentum of a lepton in $J/\psi \rightarrow l^+ l^-$ 
and the three--momentum of a pion in $\rho \rightarrow \pi \pi$ decay.

%Another signature for this possibility is to look for $X \rightarrow (3 \pi)
%J/\psi$, which is just below the $X \rightarrow \omega J/\psi$. 
%If $\omega J/\psi$ were kinematically allowed, its branching ratio should be
%the same as $X\rightarrow \rho J/\psi$, since $X$ is  made from
%$b\rightarrow c \bar{c} s \rightarrow (c \bar{c} d \bar{d} ) s$.
%Since the final $DD$ state is in $I=0$ and $I=1$ with equal strength, 
%$X\rightarrow \rho J/\psi$ and $X\rightarrow \omega J/\psi$ should have
%the same branching ratio. Since the available energy is slightly below
%the $\omega J/\psi$ threshold, the rate should be suppressed. But still
%$X \rightarrow (3 \pi) J/\psi$ could be noticable.

\section{conclusion}

In this letter, we pointed out that the dipion invariant mass spectrum in
$X (3872) \rightarrow J/\psi \pi \pi$ could be useful in determination of
$J^{PC}$ quantum number of the newly observed resonance $X (3872)$.
In particular, the current preliminary data seems to already exclude the
possibility $X = {^1P}_1$, since the dipion invariant mass spectrum has 
a peak at low $m_{\pi\pi}$ region in this case.
This is apparently disfavored by the preliminary Belle data. 
On the other hand, if $X = {^3D}_2 (2^{--})$ or $X = {^3D}_3 (3^{--})$, 
then the $m_{\pi\pi}$ spectrum has a peak at high $m_{\pi\pi}$ region, 
which is consistent with the data. 
Also the interpretation in terms of $D \overline{D^*}$ molecular state 
predicts $m_{\pi\pi}$ spectrum to have a sharp peak near 
$m_{\pi\pi} \approx m_\rho$. This could be easily distinguishable in the
future when more data are accumulated. 

\begin{acknowledgments}
I thank T. Barnes, G. Bodwin, Sookyung Choi, E. Eichten, Jungil Lee,
S. Olsen and M.B. Voloshin for useful communications and discussions.
This work is supported in part by BK21 Haeksim Program and 
KOSEF through CHEP at Kyungpook National University.
\end{acknowledgments}

\end{document}